\documentstyle[aaspp4,12pt,epsfig,color]{article}
\begin{document}

\title{A Look At Three Different Scenarios for Bulge Formation}

\author{Rychard Bouwens}
\affil{Physics Department,
   University of California,
    Berkeley, CA 94720; bouwens@astro.berkeley.edu}
\centerline \&
\author{Laura Cay\'on}
\affil{Instituto de F\'\i sica de Cantabria,
       CSIC-Universidad de Cantabria, 39005 Santander (Cantabria),
       Spain; cayon@ifca.unican.es}
\centerline \&
\author {Joseph Silk}
\affil{Astronomy and Physics Departments, and Center for Particle
    Astrophysics, University of California,
     Berkeley, CA 94720; silk@astro.berkeley.edu}

\begin{abstract}
In this paper, we present three qualitatively different scenarios for
bulge formation: a secular evolution model in which bulges form after
disks and undergo several central starbursts, a primordial collapse
model in which bulges and disks form simultaneously, and an
early bulge formation model in which bulges form prior to disks.  We normalize
our models to the local $z=0$ observations of de Jong \& van der Kruit
(1994) and Peletier \& Balcells (1996) and make comparisons with high
redshift observations.  We consider model predictions relating
directly to bulge-to-disk properties.  As expected, smaller
bulge-to-disk ratios and bluer bulge colors are predicted by the
secular evolution model at all redshifts, although uncertainties in
the data are currently too large to differentiate strongly between the
models.

\end{abstract}


\section{Introduction}
A number of different mechanisms have been proposed for the formation
of bulges: primordial collapse (Eggen, Lynden-Bell, \& Sandage 1962),
hierarchical galaxy formation models (Kauffmann \& White 1993, Baugh
et al. 1996), infall of satellite galaxies, and the secular evolution
of galaxy disks.  Numerous arguments have been put forward that
secular evolution of disks has occurred in at least some galaxies,
particularly in late-type galaxies (Kormendy 1992; Courteau 1996).
However, for some galaxies, notably those with a massive bulge, simple
energy arguments show that not all galaxies could have formed in this
way.  Such galaxies would have necessarily formed by
primordial collapse, major mergers at high redshifts, or infall of
satellite galaxies (Pfenniger 1992).  In summary, it appears that many
mechanisms have been at work in forming bulges over the history of
universe, and so the question is no longer which mechanisms were
effective in forming bulges but in what fraction.

In this paper, we shall broadly classify these bulge formation
scenarios into three types: secular evolution in which bulges form
relatively late by a series of bar-induced starbursts, one in which
bulges form simultaneously with disks, and an early bulge formation
model in which bulges form earlier than disks.  Adjusting the three
models to produce optimal agreement with $z=0$ observations, we
compare their high-redshift predictions with present observations, in
particular, with data compiled in various studies based on the CFRS
(Schade et al.\ 1995; Schade et al.\ 1996; Lilly et al.\ 1998) and the
HDF (Abraham et al.\ 1998).

We begin by presenting the samples used to constrain the models (\S2),
follow with a description of the models (\S3), provide a brief
description of our computational method (\S4), move onto our
high-redshift predictions and comparisons with available observations
(\S5), and finally summarize the implications of our analysis (\S6).
Hereinafter, we use $H_o = 50$ km/s/Mpc.

\section{Local $z=0$ Samples}

For the purposes of normalizing our models, we examine two local $z=0$
samples: the de Jong sample (de Jong \& van der Kruit 1994; de Jong
1995, 1996; hereinafter, DJ) and the Peletier \& Balcells sample of
galaxies (Peletier et al.\ 1994; Peletier \& Balcells 1996;
hereinafter, PB).  The DJ sample is selected from $\sim12.5\%$ of the
sky and considers only relatively face-on ($b > 0.625$) galaxies
(37.5\% of all orientations).  For simplicity, we shall treat
selection of this sample to be on $\sim4.7\%$ of the sky ({\it e.g.,}
0.59 steradians).  Following de Jong (1996), we also take it to be
diameter-limited in $R$ to galaxies larger than $2'$ at 24.7 R-mag
$\textrm{arcsec}^2$.

The PB sample is a similarly diameter-limited sample: the $B$-band
diameter in terms of its $25$ mag/arcsec$^2$ isophote was
restricted to the range $90''$ and $150''$.  However, in contrast to
the relatively face-on ($b > 0.625$) DJ sample, the PB sample
considers galaxies of all orientations, and this was our principal
reason for including it in our comparisons.  Unfortunately, the PB
sample is more restricted than the DJ sample in the Hubble types
included (3.0 to 6.5) and in the surface brightness range $(20.5 <
\mu_0 ^{b_J} < 21.5)$.

In the model comparisons which follow, we select our local $z=0$
subsample from the local samples using the DJ selection criteria since
the PB sample is roughly a subset of the DJ sample strictly in terms
of the selection criteria.  We attempt to normalize the PB sample
relative to the DJ sample so that it contains 31\% of the number in
the de Jong sample since $\sim 752$ out of 1207 galaxies (62\%) in the
ESO-LF catalogue down to $1'$ (Lauberts \& Valentijn 1989) were of
type 3.0 to 6.5 (Sbcd) and roughly 50\% of the DJ sample was in the PB
surface brightness selection range.  In principle, then, the PB sample
should be a simple subset of the DJ sample.  Unfortunately, a simple
look at the relative colour and $B/T$ distributions for the DJ and PB
samples indicates that there are more galaxies in the PB sample with
large $B/T$ ratios and relatively blue bulges than in the DJ sample.
Many of these differences can be attributed to the fact that the
properties of the DJ sample were measured from face-on galaxies while
the PB sample covered a range of inclination angles.  Edge-on disks in
the PB sample are simply redder and less prominent relative to the
bulges due to the greater path length the light must traverse through
the dusty disks.

In all of the model comparisons which follow, due to the various
complications associated with the exact meaning of UGC $R$ diameter,
the relative fraction of low surface brightness galaxies, and the
influence of disk orientation on selection, we simply adjust the
surface brightness threshold at which the UGC diameters (from which
the DJ sample was taken) were measured to obtain rough agreement with
the number of galaxies obtained in the DJ sample.

\section{Models}

Starting with the local properties of disks and a reasonable
distribution of formation times, we construct a fiducial disk
evolution model, to which we add three different models for bulge
formation, the principle difference being simply the time the bulges
form relative to that of their associated disks.  Since it is simply
our intent to examine the extent to which current observations allow
us to discriminate the order in which bulges and disks form, we
intentionally do not consider a more complex model (e.g., Kauffmann et
al.\ 1993; Baugh et al.\ 1996; Molla \& Ferrini 1995) nor do we
attempt to model the internal dynamics or structure of spirals (e.g.,
Friedli \& Benz 1995).

We assume the Sabc and Sdm luminosity functions (LFs) for disk
galaxies given by Binggeli, Sandage, and Tammann (1988).  We adjust
the bulge-to-total ($B/T$) distributions of these galaxy types to
obtain fair agreement with those distributions measured in the DJ and
PB samples (see Figure 3).  We evolve these galaxies backwards in time
in luminosity according to their individual star formation histories
without number evolution, presuming that significant evolution in
number occurs only at redshifts above those examined in the present
study ($0 < z < 1$).  For this reason, we do not make predictions
above $z \sim 1$.

We take the formation times of these galaxies to be distributed
identically to that given by the procedure outlined in Section 2.5.2
of Lacey \& Cole (1993) except that we take halo formation time to
equal the time over which 0.25 of the final halo mass is assembled.
For the purposes of calculating halo formation times corresponding to
galaxies of a given luminosity, we assume a constant mass-to-light
ratio where a $M_{b_J} = -21.1$ galaxy has $4 \cdot 10^{12} M_{\odot}$
and we adopt the CDM matter power spectrum given in White \& Frenk
(1991):
\begin{equation}
P(k) = \frac{1.94 \times 10^4 b^{-2} k}
{(1 + 6.8 k + 72 k^{3/2} + 16 k^2)^2} \textrm{Mpc}^3
\end{equation}
For $b = 1$, this expression yields $\sigma_8 = 1.$  

Just as we choose to take the halo formation time to be the time over
which 0.25 of the final halo mass is assembled instead of 0.5 used by
Lacey \& Cole (1993), we choose $\Omega = 0.15$ to push the epoch of
large-scale merging to high enough redshift so that the observed
number of stars are able to build up in these galaxies without being
destroyed by the merging events prevalent at earlier epochs.  Since we
have observable constraints on the star formation history of the
universe, there is a certain epoch after which disks must remain
largely undisturbed.  Of course, if we had assumed that some fraction
of the stars in the disk were added by minor mergers, we could push
the halo formation time, and consequently the formation of disks and
bulges, to lower redshift by raising the value of $\Omega$.  We
illustrate the distribution of halo formation times in Figure 1 for
several different luminosity ranges.

We take star formation in the disk to commence at the halo formation
time with an e-folding time that depends on the $z=0$ galaxy
luminosity, i.e., $\tau = (3\,\textrm{Gyr}) 10^{0.4(M_{b_J} + 20)}$ to
roughly fit the $z=0$ colour-magnitude relationship (see Figure 2), so
that the star formation rate in the disk of a galaxy with absolute
magnitude $M_{b_J}$ and halo formation time $t_{HF}$ can be expressed:
\begin{equation}
SFR_{disk} \propto \left\{
\begin{array}{ll} e^{-(t_{HF}-t)/{\tau}} & t < t_{HF},\\
0 & t \geq t_{HF}.
\end{array}
\right.
\end{equation}
We adopt the standard equations for evolution in metallicity to $z=0$
(Tinsley 1980) and tune the yields for each luminosity separately to
reproduce the $z=0$ disk metallicities given by
$[Fe/H]=(-0.17)(M_Z+20)-0.28$ (Zaritsky, Kennicutt, \& Huchra 1994).
Since we are not trying to develop a universal model for chemical
evolution in disks, we have simply tuned the yields for each
luminosity separately.

Using the Calzetti (1997) extinction prescription and a screen
model for the dust, we take the optical depth $\tau$ of dust in the
$B$ band to equal $0.7 (10^{-0.17 (1.3) (M_{B} + 19.72)})$, consistent
with the values given in Peletier et al.\ (1995).  We assume
exponential profiles for the disks with a $b_J$ central surface
brightness given by $21.65 + 0.2 (M_B + 21)$ for simplicity where this
expression accounts for the observed correlation between surface brightness and
luminosity  (e.g., de Jong 1996; McGaugh \& de Blok 1997).
We compute spectra for the purposes of determining colours and
magnitudes using the Bruzual \& Charlot instantaneous-burst
metallicity-dependent spectral synthesis tables as compiled in
Leitherer et al.\ (1996).  For metallicities in between those compiled
here, we have interpolated between the provided tables in units of
$\log Z$.

To calibrate our fiducial disk evolution models, we compare the model
predictions to both the colour-magnitude relationship of disks in
spirals and the cosmic history of luminosity density.  Firstly, with
regard to the colour-magnitude relationship, we note that we produce
good agreement with the colour-magnitude relationship given in the DJ
and PB samples, both in terms of their slopes and overall
distributions (Figure 2).  Given our relatively reasonable assumptions
about the quantity of dust and metals in these galaxies, matching
these distributions gives us a basic constraint on the star formation
history in disk galaxies of different luminosities.  Secondly, all
models, for which bulge, disk, and E/S0 contributions have been
considered, produce fair agreement with the luminosity density of the
universe at all redshifts for which observable constraints are
available (Lilly et al.\ 1996; Madau et al.\ 1996; Connolly et al.\
1997), though the observed luminosity density is slightly lower at
lower redshifts (Figure 3).  Resolving this discrepancy requires
pushing the formation of disk galaxies to higher redshifts, i.e.,
lowering $\Omega$.  Discrepancies in the ultraviolet luminosity
density could be easily removed by introducing moderate dust
extinction at high $z$, as motivated by many recent analyses (Sawicki \&
Yee 1998; Calzetti 1997; Meurer et al.\ 1997).  Note that the
similarities of the models at high redshift follows from the dominant
and identical E/S0 contribution.  Having described our fiducial
disk model, we now describe the basic three models for bulge formation
that we will be comparing.

\noindent{\it Secular Evolution Model:} In the secular evolution
scenario, bulges form after disks.  In this scenario, gas accretion
onto the disk triggers the formation of a bar, gas-inflow into the
center, and then star formation in the galaxy center (Friedli \& Benz
1995; Norman, Sellwood, \& Hasan 1996).  The build-up of a central
mass destroys the bar and inhibits gas inflow, consequently stopping
star formation in the bulge until enough gas accretes onto the galaxy
to trigger the formation of a second bar, gas inflow into the center,
and finally a second central starburst.  Somewhat arbitrarily, we
suppose that the first central starburst occurs some 2 Gyr after disk
formation in our fiducial model, that central starbursts last 0.1 Gyr,
a time-scale matching those found in the detailed simulations by
Friedli \& Benz (1995), and that 2.4 Gyr separates central starbursts,
numbers used just to illustrate the general effect of a late secular
evolution model for the bulge.  We repeat this cycle indefinitely and
assume that the star formation rate follows an envelope with an
e-folding time equivalent to the history of disk star formation:
\begin{equation}
SFR_{bulge} \propto \left\{
\begin{array}{ll} e^{-(t_{HF}-2 Gyr -t)/{\tau}} & \frac{t_{HF} - 2 
Gyr - t}{2.5 Gyr} - \lfloor \frac{t_{HF} - 2 
Gyr - t}{2.5 Gyr} \rfloor < 0.04,\\ 0 & \frac{t_{HF} -2 Gyr -
t}{2.5 Gyr} - \lfloor \frac{t_{HF} -2 Gyr -
t}{2.5 Gyr} \rfloor \geq 0.04,\\ 0 & t \geq t_{HF} - 2 Gyr
\end{array}
\right.
\end{equation}
where $\lfloor \,\,\rfloor$ is the greatest integer function.  We thereby
force star formation in the disk and the bulge to follow very similar
time scales, given the extent to which they are both driven by gas
infall processes.  Of course, bulge growth over the history of the
universe should affect these time-scales, but given the already large
uncertainties in both gas accretion and star formation, we have
decided to ignore this.  For all bulge models, we adopt the slope of
the approximate luminosity-metallicity relationship
$[Fe/H]=-(0.02/0.135)M_R - 3.1852$ (Gonz\'alez \& Gorgas 1996;
Jablonka et al. 1996; Buzzoni et al. 1992).  For the secular evolution
model we fix the metallicity at the $z=0$ value somewhat crudely to
account for the fact that this gas would already be polluted by stars
which formed in the disk.

\noindent{\it Simultaneous Formation Model:} We assume for our second
model that star formation in the bulge commences at the formation time
of disks in our fiducial model.  In this model, high angular momentum
gas forms the disk while the low angular momentum gas simultaneously
forms the bulge.  As in the secular evolution model, we suppose that
the star formation in the bulge lasts $\tau_{burst}$ = 0.1 Gyr so that
\begin{equation}
SFR_{bulge} \propto \left\{
\begin{array}{ll} e^{-(t_{HF}-t)/{\tau_{burst}}} & t < t_{HF},\\
0 & t \geq t_{HF}.
\end{array}
\right.
\end{equation}
To obtain distributions of bulge colours for both the simultaneous and
the early bulge formation models that match the data, we
systematically decrease the metallicity of bulges by 0.2 relative to
the relationship preferred by Jablonka et al.\ (1996).  As in the
disk, we assume evolution in the metallicity of the gas that forms the
bulge.

\noindent{\it Early Bulge Formation Model:} In models where bulges
form through the merging of disk galaxies, the formation of the stars
found in bulges is expected to precede the formation of stars in the
disks which form out of gas which accretes around the spheroid (e.g.,
Kauffmann \& White 1993; Frenk et al.\ 1996).  For simplicity, we
commence star formation in the bulge 4 Gyr prior to the formation of
disks in our fiducial model and suppose that it lasts $\tau_{burst} =
0.1$ Gyr as in our other models so that
\begin{equation}
SFR_{bulge} \propto \left\{
\begin{array}{ll} e^{-(t_{HF} + 4 Gyr - t)/{\tau_{burst}}} & t < t_{HF} + 4 
Gyr,\\
0 & t \geq t_{HF} + 4 Gyr.
\end{array}
\right.
\end{equation}

Finally, to these models, we add a simple model for E/S0 galaxies to
aid with the interpretation of observed high redshift, high $B/T$
systems.  We adopt the same luminosity function for the E/S0 galaxies
given in Pozzetti et al.\ (1996) but with a $20\%$ higher
normalization to somewhat better fit the observed evolution in
luminosity densities.  We somewhat arbitrarily assume that the
distribution of formation redshifts for the $E/S0$ population is
scaled to be at exactly twice the distribution of formation redshifts
obtained for the same luminosity spiral in the $b_J$ band so that if
the median formation redshift for some luminosity spiral is 1, the
median formation redshift for the same luminosity E/S0 is 2.  We take
the e-folding time for star formation to be 0.5 Gyr.  Since it is
likely that the stars in ellipticals were assembled from other
galactic fragments through mergers, this scenario is only intended to
be representative of when the stars in elliptical galaxies formed
rather than where they formed.  We assume that the $E/S0$ population
has $B/T$ ratios distributed between 0.5 and 1 with a scatter intended
to represent both the intrinsic uncertainty in the relative local mix
of $E$ and $S0$ galaxies and the realistic scatter in $B/T$ values
extracted in typical bulge-to-total luminosity decompositions ({\it
e.g.}, Ratnatunga et al.\ 1998).  We further assume that the
metallicity of all the stars in our E/S0 population are of solar
metallicity.

\section{Computational Method}

We perform the calculations by considering four different
morphological types, dividing each type into 3 different luminosity
classes (where the width of each class in absolute magnitude is 2),
and allowing each of the luminosity classes for a specific type to
form at 20 different discrete redshift intervals from $z=0$ to
$z=2.5$, the relative proportion being determined by the distribution
of formation times for galaxies of a specific luminosity (as discussed
in \S3).  Then, for each of these $4 \times 3 \times 20 = 240$
distinct galaxy evolution histories, we determine how the gas,
metallicity, star formation rate, and luminosity evolves.  Finally,
Monte Carlo catalogues are constructed using the quoted selection
criteria, the given number densities, and these computed luminosity
histories.

\section{High  Redshift Comparisons}

As already mentioned, all our models have been constrained to
reproduce the bulge-to-total distribution for the DJ and PB samples as
can be seen in Figure 4.  Comparisons with bulge $B-R$ colors and
differences between bulge and disk $B-R$ colors are presented in
Figure 5.  Both our early bulge formation model and simultaneous bulge
formation model produce good fits to the bulge colours and relative
bulge-to-total colours at low redshift.  Clearly, in the secular
evolution model, not only are the bulges of local galaxies too blue
relative to the disks, but there is more dispersion in both the bulge
colours and bulge-to-disk relative colours than there is in low
redshift samples.  If necessary, inclusion of a small amount of
reddening in the models would give better agreement with the low
redshift data.  It is also possible that the irregular morphology
and/or potential AGN activity might cause these blue-bulged galaxies
to be removed from the local samples.

We now examine the predictions of these different scenarios in terms
of the higher redshift ($z\sim1$) observations.  Since the essential
difference between the models is the formation time of bulges relative
to the fiducial formation time of disks, we shall focus on the
observables directly contrasting the bulge and disk properties: in
particular, the high-redshift bulge-to-disk ratios and the
high-redshift bulge-to-disk colours in our comparisons (see Figures
4-6).  We begin by comparing our models to the bulge-to-total ratios
of high redshift galaxies in the CFRS sample in Figure 4.  We examine
both the ground-based sample of Schade et al.\ (1996) and the
HST-selected large disk ($r > 4h_{50}$ kpc) subsample of Lilly et al.\
(1998) in three different redshift intervals.  To compare our models
with the observations, we have generated Monte-Carlo catalogues of
galaxies with bulge-to-total ratios in the observed $F814W$ band,
applying the CFRS selection criteria ($I_{AB} < 22.5$ and a central
$I_{AB}$ surface brightness $< 24.5$) and the size cut ($h > 4$ kpc)
to compare specifically with the Lilly et al.\ (1998) large disk
subsample.

We subdivide the samples into the redshift bins (0.2,0.5), (0.5,
0.75), and (0.75, 1.00).  While the differences between the models are
quite small at low $z$, interesting differences begin to arise at
$z\sim1$.  Unfortunately, at $z\sim1$, the observed $F814W$ band is
approximately probing rest-frame $B$ light and hence is quite
sensitive to active star formation.  Consequently, the ordering of
Models I, II, and III in terms of the number of galaxies with large
B/T ratios is not the same as the order in which bulges form in these
three models.  The secular evolution model (Model I), with late bulge
formation, has a paucity of large B/T objects relative to the other
models.  The simultaneous bulge formation model (Model II) has a large
number of such galaxies simply because a large number of bulges were
forming at this time, while the early bulge formation model (Model
III) has a slightly lower value due to the fact that bulges in this
model had long been in place within their spiral hosts.  Presumably,
high resolution infrared images as will be available with NICMOS
should be a more powerful discriminant between these models since it
is more sensitive to total stellar mass than it is to current star
formation.  Unfortunately, both the lack of data and uncertainties in
this data ($\pm 0.2$ in B/T) (Schade et al.\ 1996) are too large to
permit any strong statements.  It does appear, nevertheless, that
there are too many large B/T systems observed (Lilly et al.\ 1998)
relative to the models, and therefore there may be a lack of high B/T
galaxies in our model of high luminosity, large disks.

We now look at the bulge colours and relative bulge-disk colours of
high redshift galaxies.  For our first sample (32 galaxies), for which
HST images of CFRS-selected galaxies were available, we utilize the
colours and bulge-to-total ratios compiled in Table 1 and Figure 5 of
Schade et al.\ (1996).  Following Schade et al.\ (1996) in the use of
the best-fitting CWW SED templates for the purposes of k-corrections
and colour conversions, we calculate the colours for the bulge-disk
components from the total integrated $(U-V)_{0,AB}$ colours, the
tabulated bulge-to-disk ratios given in the rest-frame $B$ band, and
the $(U-V)_{0,AB}$ colours of the indicated component.  For our second
sample (27 galaxies), we consider the bulge-to-disk colours compiled
by Abraham et al.\ (1998) from a subsample of the Bouwens, Broadhurst,
\& Silk (1998a) sample, for which both $z>0.3$ and fits to the
bulge-to-total ratio were available (Ratnatunga et al.\ 1998).  Note
that the bulge (disk) colours compiled by Abraham et al.\ (1998) are
determined from the light inside (outside) a 3-pixel aperture and are
not determined from a proper bulge-disk decomposition.  Using the
best-fit CWW SED templates, we convert the Abraham et al.\ (1998)
colours to their rest-frame values.  We plot the data separately for
these two samples due to potentially different systematics.  Because
of the larger uncertainties involved in determining the relative
bulge-disk colours for galaxies dominated by a bulge ($B/T > 0.55$) or
disk component ($B/T < 0.1$), we have excluded these galaxies from our
comparisons due to the potentially large errors in the determination
of the disk and bulge colours separately.  For both data sets, we
again compare with Monte-Carlo catalogues generated using the CFRS
selection criteria due to its close similarity with the Bouwens et
al.\ (1998a) selection criteria ($I_{F814W,AB} < 22.33$).  We present
histograms of the bulge colours and relative bulge-to-disk colours in
Figure 5, and a scatter plot of the bulge-to-total ratios both versus
the bulge colours and versus the relative bulge-to-disk colours in
Figure 6.  For both Figures 5-6, we subdivide the galaxies into the
redshifts bins (0.3, 0.5), (0.5, 0.75), and (0.75, 1.0).

As expected, in all redshift bins, bulges are slightly bluer in the
late bulge formation models than are the disks (Figure 5).  A blue
tail may be marginally detectable in the Schade {\it et al.} data in
the highest redshift bins.  Unfortunately, given the extremely limited
amount of data and uncertainties therein, little can be said about the
comparison of the models in all three redshift bins, except that the
range of bulge and relative bulge-to-disk colours found in the data
appears to be consistent with that found in the models.

Figure 6 shows that the scatter in the data can readily be reproduced
at both low and high redshift for the various models.  Clearly, the
secular evolution and other bulge formation models separate out in
this diagram, late bulge formation models always yielding bluer bulges
for a given B/T ratio.  Unfortunately, the observational data set is
sufficiently small and contains enough uncertainties (an estimated
$\pm 0.1$ in the B/T ratio and $\pm 0.3$ in relative bulge-disk
colours) that it is difficult to verify whether there is a paucity of
blue bulges at high redshift relative to the predictions of the
secular evolution model, though there appears to be several bluer
bulges in the redshift interval (0.5,0.75).

\section{Summary}

We have developed three representative models for bulge formation and
evolution.  While consistent with currently available data, our models
are schematic and are intended to illustrate the observable
predictions that will eventually be made when improved data sets are
available in the near future.  Our models are (i) secular evolution,
in which disks form first, (ii) simultaneous formation of bulge and
disk, as might be expected in a monolithic model, and (iii)
early bulge formation, in which bulges form first.  We normalize to
two local $z=0$ samples which provide template bulge and disk
luminosity ratios and colours.  We make predictions for these bulge
and disk parameters to $z\sim1$ for comparison with observed samples.

Admittedly, our models are still quite crude, assuming among other
things that the effects of number evolution on the present population
of disks can be ignored to $z \sim 1$ as suggested, for example, in
Lilly {\it et al.} 1998.  Of course, one recent analysis (Mao, Mo, \&
White 1998) has argued that observations favor the interpretation that
a non-negligible amount of merging has taken place in the disk
population from $z=0$ to $z=1$.  For this particular interpretation,
it is not clear to us how all the present stellar mass in disks could
have built up if disks were continually destroyed by merging to low
$z$ given the constraints on the cosmic star formation history.

We have also not considered the environmental dependencies that are
sure to be important in the generation of the Hubble sequence.  We
plan on addressing these shortcomings in future work (Bouwens et al.\
1998b) in the context of a semi-analytical hierarchical clustering
model where we consider the formation of bulges by both secular and
hierarchical evolution.

\acknowledgements

We acknowledge useful discussions with Roberto Abraham, Marc Balcells,
Francoise Combes, Roelof de Jong, David Friedli, Kavan Ratnatunga, and
David Schade.  This document has also been improved based upon
suggestions by the scientific editor Steven Shore and an anonymous
referee.  RJB is grateful for support from an NSF graduate fellowship
and JS from NSF and NASA grants as well as support as from the Chaire
Blaise Pascal.  LC thanks the Astronomy Department and the CfPA
(Berkeley) as well as the IAP (Paris) for their hospitality during her
stays in those institutes.  LC acknowledges support from the Spanish
DGES PB95-0041.  The Medium Deep Survey catalog is based on
observations with the NASA/ESA Hubble Space Telescope, obtained at the
Space Telescope Science Institute, which is operated by the
Association of Universities for Research in Astronomy, Inc., under
NASA contract NAS5-26555. The Medium-Deep Survey was funded by STScI
grant GO2684.

{}

\newpage

\begin{figure}
\epsscale{0.95}
\plotone{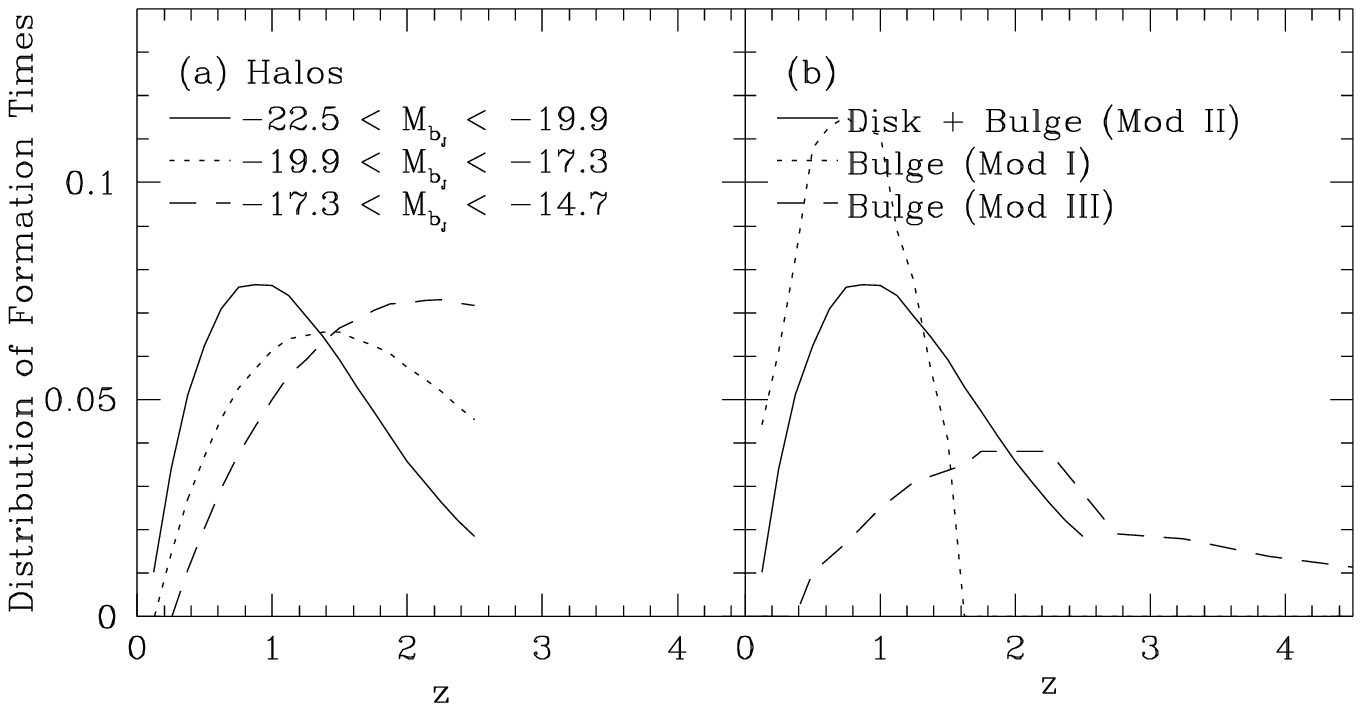}
\caption{(a) Distribution of halo formation times for disk galaxies of
different luminosities.  (b) Distribution of disk and bulge formation
times in Models I (secular evolution model), II (simultaneous
formation model), and III (early bulge formation model).}
\end{figure}

\newpage

\begin{figure}
\epsscale{0.95}
\plotone{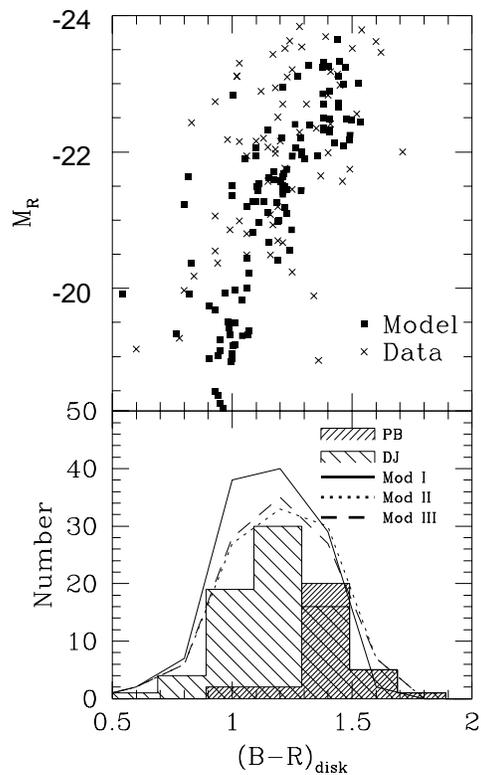}
\caption{Comparison of local disk properties with our model
predictions.  The lower panel presents the $B-R$ colour distribution.
Mod I (solid line) corresponds with the secular evolution model, Mod
II (dotted line) corresponds with simultaneous formation model, and
Mod III (short dashed line) corresponds with the early bulge formation
model.  The upper panel presents a scatter plot of the observed and
predicted colour-magnitude relationship.  For both panels,
observational data are taken from Peletier \& Balcells 1996 (dashed
histogram) and de Jong 1995 (solid line histogram).  Note that all the
models use exactly the same disk model and thus give the same results,
except for small differences observable in the figure due to our use
of three different realizations.}
\end{figure}

\newpage

\begin{figure}
\epsscale{0.95}
\plotone{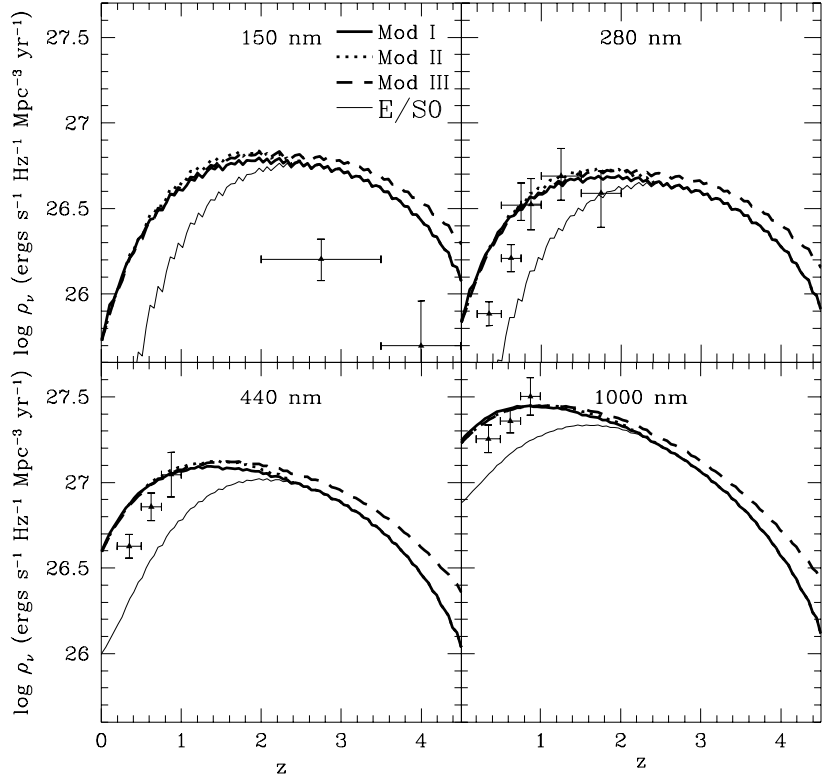}
\caption{Comparison of the observed luminosity density at 150 nm,
280 nm, 440 nm, and 1000 nm against that predicted from our models
including both the disk, bulge, and elliptical components.  The
contribution of E/S0 galaxies has been explicitly plotted to
illustrate its contribution to the total.
Observational data is taken
from Madau et al.\ (1996) at 150 nm, Lilly et al.\ (1996) and Connolly
et al.\ (1997) at 280 nm, Lilly et al.\ (1996) at 440 nm, and Lilly et
al.\ (1996) at 1000 nm.}
\end{figure}

\newpage

\begin{figure}
\epsscale{0.95}
\plotone{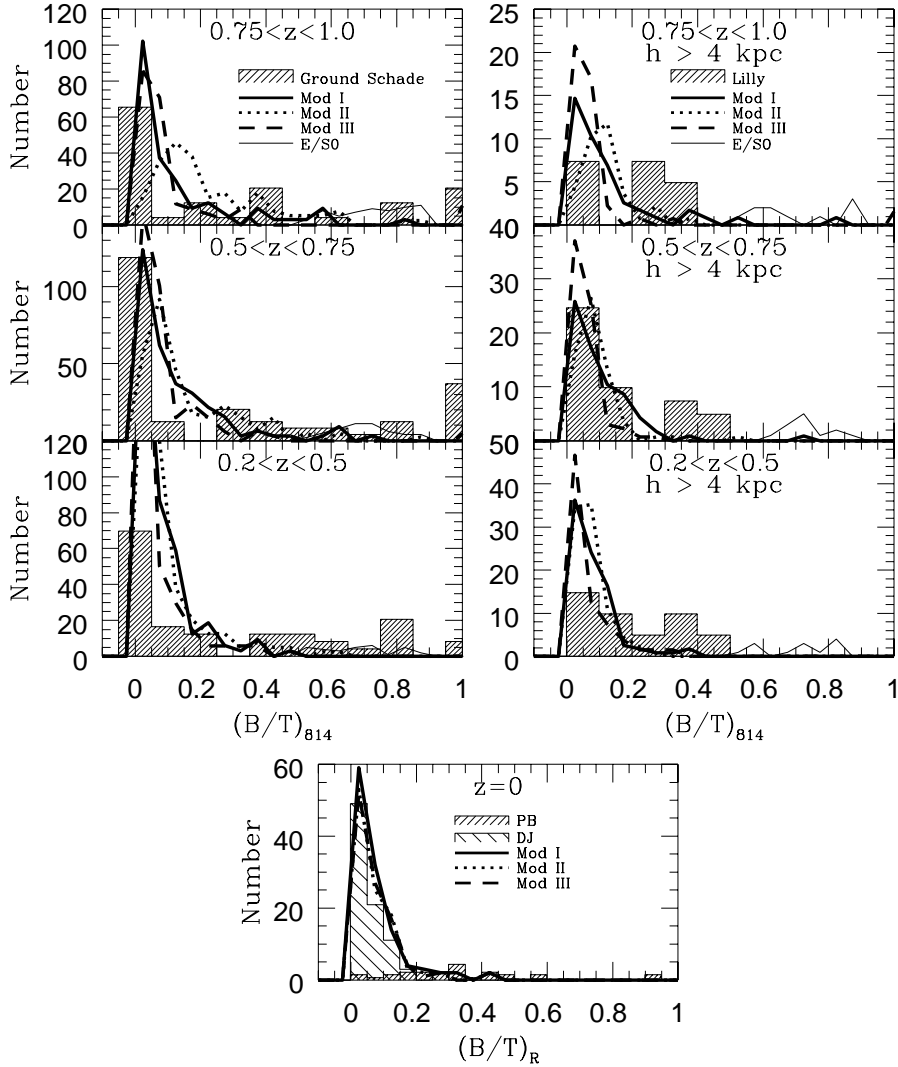}
\caption{Comparison of the observed bulge-to-total ratios (histograms)
with three different bulge formation models (see Figure 2 for a
description) at both high and low redshift.  High redshift comparisons
are performed in the upper left panels against the Schade et al.\
(1996) data using the CFRS selection criteria and in the upper right
panels against the Lilly et al.\ (1996) data using the CFRS selection
criteria plus a size cut ($h > 4 kpc$).  $E/S0$ predictions are also
included in the high redshift figures (long dashed line).  Models are
renormalized to match the data.  Data used for the low-redshift
comparisons are as in Figure 2.}
\end{figure}

\newpage

\begin{figure}
\epsscale{0.95}
\plotone{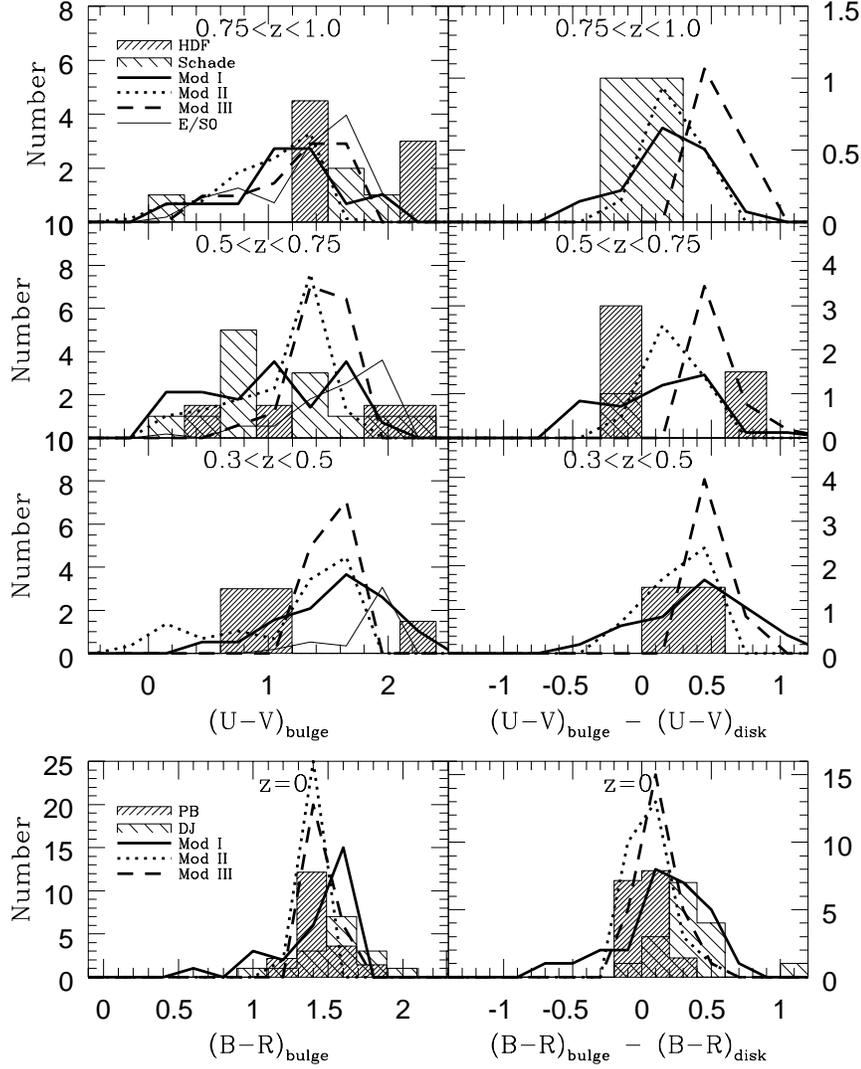}
\caption{Comparison of the observed bulge and relative bulge-to-disk
colours (histograms) with those of the models, at both high and low
redshift.  Model curves (renormalized to match observations and
multiplied by 1.6 to increase their prominence) and low redshift data
are represented as in Figure 4.  The high redshift comparison includes
data from the HDF for the Bouwens et al.\ (1998) sample (shaded
histogram) and HST data from Schade et al.\ (1995) (open histogram).}
\end{figure}

\newpage

\begin{figure}
\epsscale{0.95}
\plotone{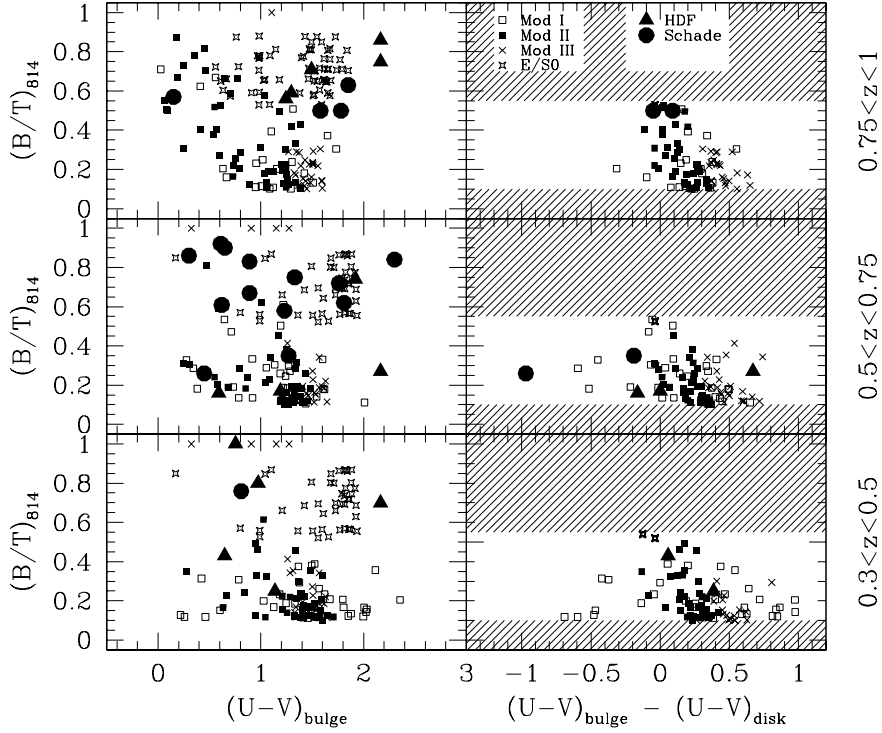}
\caption{Observed bulge-total ratio vs. bulge and relative
bulge-to-disk colours compared to Monte-Carlo realizations of these
same quantities for our secular evolution model (open squares), our
primordial collapse model (solid squares) and our early bulge
formation model (crosses) as well as our E/S0 model (stars). The
observational data are taken from the same two samples as in Figure 5,
the Bouwens et al.\ (1998) HDF sample (solid triangles) and the Schade
et al.\ (1996) sample (solid circles). The shaded areas in the right
panels correspond to galaxies dominated by the bulge ($B/T > 0.55$)
and by the disk ($B/T < 0.1$) excluded from the comparison for reasons
that are explained in the text.}
\end{figure}

\end{document}